\documentclass[aps,prl,twocolumn,floatfix,superscriptaddress]{revtex4}

\usepackage{graphicx}
\usepackage{amsmath}
\usepackage{amsfonts}
\usepackage{amssymb}

\begin{document}
\bibliographystyle{unsrt}

\newcommand{\norm}[1]{\ensuremath{| #1 |}}
\newcommand{\aver}[1]{\ensuremath{\langle #1 \rangle}}
\newcommand{\ket}[1]{\ensuremath{| #1 \rangle}}
\newcommand{\fref}[1]{Fig.~\ref{#1}}

\newcommand{\figwidth}{0.8\columnwidth}

\title{Phase diagram of hole doped two-leg $Cu$-$O$ ladders}
\author{P. Chudzinski}
\affiliation{Laboratoire de Physique des Solides, Bat. 510, Universit\'e Paris-Sud 11, Centre d'Orsay, 91405 Orsay Cedex, France}
\author{M. Gabay}
\affiliation{Laboratoire de Physique des Solides, Bat. 510, Universit\'e Paris-Sud 11, Centre d'Orsay, 91405 Orsay Cedex, France}
\author{T. Giamarchi}
\affiliation{DPMC-MaNEP, University of Geneva, 24 Quai Ernest-Ansermet CH-1211 Geneva, Switzerland }

\begin{abstract}
In the weak coupling limit, we establish the phase diagram of a two-leg ladder with a unit cell containing both $Cu$ and $O$ atoms, as a function of doping. We use bosonization and design a specific RG procedure to handle the additional degrees of freedom. Significant differences are found with the single orbital case;
for purely repulsive interactions, a completely massless quantum critical region is obtained at intermediate carrier concentrations (well inside the bands) where the ground state consists of an incommensurate pattern of orbital currents plus a spin density wave (SDW) structure.
\end{abstract}
\maketitle

The challenging physics of strongly correlated systems provides a unique opportunity to test many proposals for new, unconventional quantum states of matter. In that respect, ladders constitute a particularly interesting case \cite{dagotto_ladder_review}. These 1D systems behave quite differently from single chains. One can show for instance that, for purely repulsive interactions, they favor superconductivity in their ground state. Understanding their properties -- both experimentally \cite{nagata_ladder_single} and theoretically -- is thus interesting in its own right but also could help us gain valuable insight into the elusive physics of the two dimensional cuprate superconductors.

Most studies of ladder compounds model these systems with a single orbital per unit cell \cite{balents_2ch,schulz_2chains}.
Using a renormalization group (RG) analysis of the Hamiltonian expressed in bosonic variables, a phase diagram can be derived in the weak coupling limit. As pointed out \cite{balents_2ch}, the parameter that may be safely tuned to arbitrary values in the weak-coupling limit is the doping $\delta$ and its variation produces a sequence of states, labelled $CnSm$, with $n$ ($m$) gapless charge (spin) modes.
For repulsive on-site Hubbard $U$ terms,
a $C1S0$ d-wave ``superconducting'' phase is found away from half filling ($\delta\neq 0$).
Relaxing the constraint on the magnitude and on the sign of the interactions and extending their range to more distant sites allows one to promote other types of orders such as orbital antiferromagnetism (OAF) \cite{orignac_2chain_long}.
This state is also
known as a flux phase and was examined in the context of the two dimensional Hubbard model \cite{affleck_marston,PALeeetal_review}.

The question of whether orbital currents could exist in
cuprate materials has received much attention. Analytical \cite{CVarma_orbitalcurrents}
and numerical \cite{srinivasan_DMRG} studies of a three band model of the copper-oxygen planes predict that, in the large $U$ limit, a strong Coulomb repulsion $V_{Cu-O}$ between
nearest neighbor $Cu$ and $O$ atoms favors such a phase. Recent experimental data
seem to support that picture \cite{BFauque_neutrons}, but more studies are
needed to confirm this scenario.

\begin{figure}
 \begin{center}
  \includegraphics[width=\figwidth]{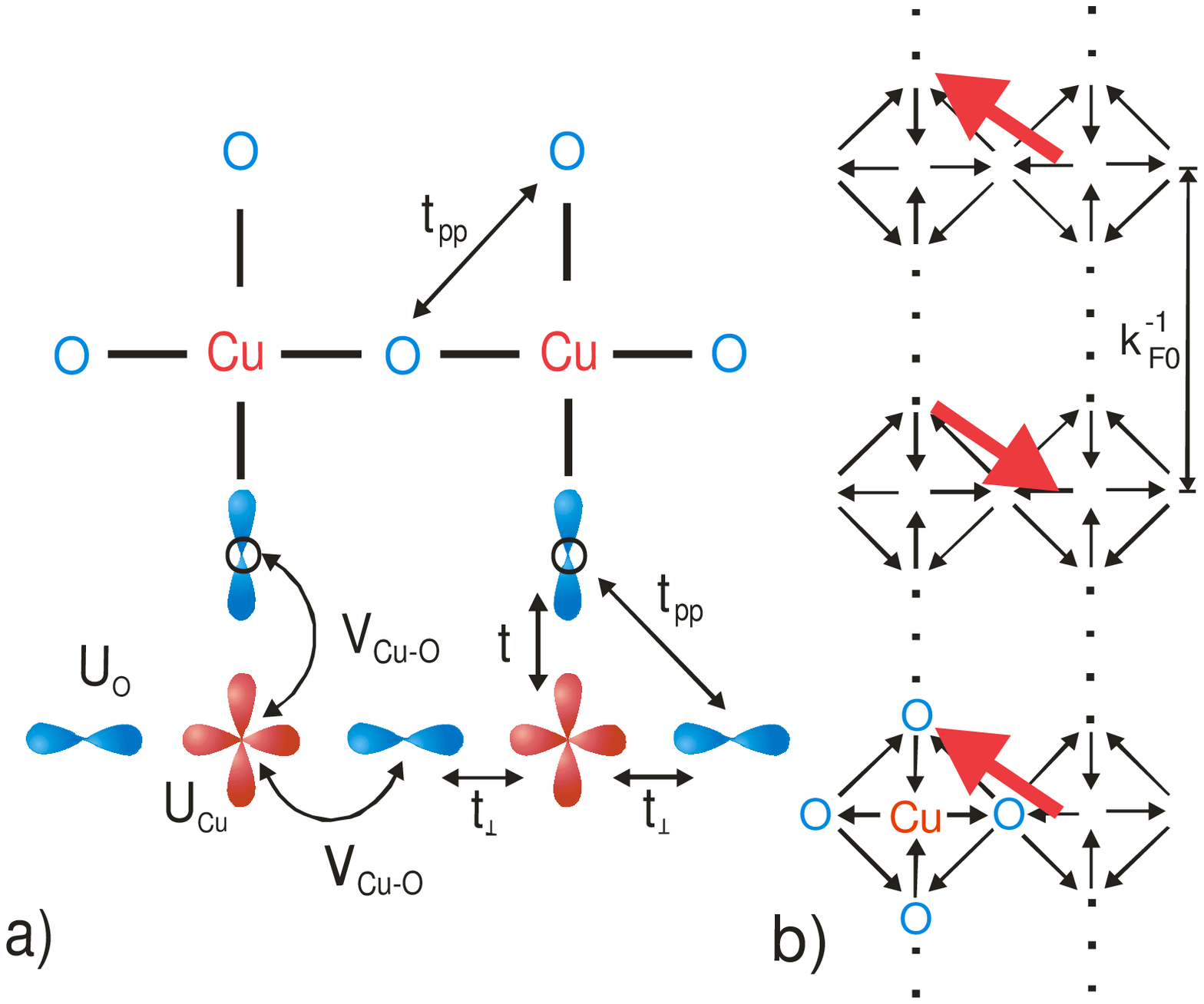}
 \end{center}
 \caption{(Color online) (a) Molecular structure of the unit cell showing the hopping and interaction parameters included in the hamiltonian (b) Orbital current pattern and
SDW modulation (bold arrows) in the $C2S2$ phase. The chain direction is vertical.
$k_{F0}$ is the Fermi wavevector in the $0$ band.
 \label{fig:structure}}
\end{figure}

We thus revisit models for ladders and include the oxygen atoms. 
Here, we focus on the issue of
whether their presence causes any significant changes
\cite{jeckelmann_DMRG} and in particular whether orbital phases
might exist for realistic choices of microscopic interactions, for
$\delta\neq 0$ \cite{lee_marston_CuO}. We establish, in the weak coupling regime,
the phase diagram as a function of hole doping of two-leg ladders
whose unit cell contains both $Cu$ and $O$ atoms, with on-site
repulsions $U_{Cu}$ ($U_{O}$) on the $Cu$ ($O$) sites and a
nearest-neighbor $V_{Cu-O}$ Coulomb term. We use RG techniques to
map out the
flows for the bosonized version of the model. 
In contrast with the case of a single orbital ladder we find that for an
intermediate range of dopings $\delta_{c1} < \delta < \delta_{c2}$
a fully massless phase is stabilized. The value of $\delta_{c1(c2)}$ depends on the bare magnitude of the Hubbard terms and/or the interoxygen hopping $t_{pp}$ which we treat as tunable parameters.
Furthermore, increasing
 $t_{pp}$ beyond a minimum value $t_{pp}^{min}$ promotes, for all
$\delta>\delta_{c1}$, an incommensurate orbital current state. It
is similar to one of the patterns advocated by Varma
\cite{CVarma_orbitalcurrents} (see Fig. \ref{fig:structure}b). The
corresponding phase has an additional SDW (CDW) character for
$\delta < \delta_{c2}$ ($\delta > \delta_{c2}$).

We consider the two leg ladder of \fref{fig:structure}a
where the relevant hopping parameters and interactions are shown.
$H_T$ is the sum of contributions describing carrier hops plus a
term proportional to $\epsilon=E_{Cu}-E_{O}$, the difference
between the $Cu$ and $O$ site energies.
Relevant values pertaining to selected copper oxide ladders have
been computed in LDA \cite{Muller-Rice_LDA}, and we use here
$t_{\bot}=1$, $\epsilon=0.5$ in units of $t$. $H_{T}$ is
diagonalized in momentum space, and since $\epsilon$ and the
various t's are of the same order, one can safely neglect the
high energy orbitals.
 We are left with two low lying bands crossing the
Fermi energy $E_F$, one bonding ($0$) and one antibonding ($\pi$)
combination of chain states. The energy dispersion is linearized
near $E_F$.
From the fermionic densities, we introduce
\cite{giamarchi_book_1d} charge (c) and spin boson (s) fields $\phi$ for each specie.
$H_{T}$ is diagonal when expressed in terms of $\phi_{\mu,\nu}$ operators
$\mu=$ c or s, $\nu = 0$ or $\pi$ in the $B_{0\pi}$ basis. The non linear terms of $H_{int}$, denoted by $H^{NL}_{int}$, have a simple form in the $B_{+-}$ basis, where $\phi_{\mu,\nu}=1/\sqrt{2}(\phi_{\mu,0}+\nu \phi_{\mu,\pi})$, $\nu = +$ or $-$.
When density-density interactions are included, the quadratic part of $H$ is diagonal in the $B_o$ basis where we introduce $\phi_{\lambda}$ operators with $\lambda=1,2,3,4$ (1,2 are s modes, 3,4 are c modes):
\begin{equation}\label{eq:Hbozon}
    H_{0}= \sum_{\lambda} \int \frac{dx}{2\pi}[(u_{\lambda}K_{\lambda})(\pi \Pi_{\lambda})^{2}+(\frac{u_{\lambda}}{K_{\lambda}})(\partial_{x} \phi_{\lambda})^{2}]
\end{equation}
The matrix $S$ which defines the rotation of the
$B_{o}$ basis with respect to $B_{+-}$ is given by
\begin{equation}
S=\frac{1}{\sqrt{2}}\left(%
\begin{array}{cccc}
  P_{1} & Q_{1} & 0 & 0 \\
  -Q_{1} & P_{1} & 0 & 0 \\
  0 & 0 & P_{2} & Q_{2} \\
  0 & 0 & -Q_{2} & P_{2} \\
\end{array}%
\right)
\end{equation}
 $P_{i}$ and $Q_{i}$ are expressed in terms of angles
$\alpha$ for the spin part and $\beta$ for the charge part;
 $P_{1 (2)}=\cos\alpha (\beta) + \sin\alpha (\beta)$ and $Q_{1 (2)}=\cos\alpha (\beta) -
\sin\alpha (\beta)$.
In the $B_{+-}$ basis $H^{NL}_{int}$ reads:
\begin{widetext}
\begin{equation}\label{eq:kosinusy}
H^{NL}_{int}=-g_{1c} \int dr \cos(2\phi_{s+})\cdot
    \cos(2\theta_{c-})+ g_{1a} \int dr \cos(2\phi_{s+})\cdot
    \cos(2\theta_{s-})- g_{2c} \int dr \cos(2\theta_{c-})\cdot
\cos(2\phi_{s-})+
\end{equation}
$$
g_{4a}\int dr\cos(2\phi_{s-})\cdot\cos(2\theta_{s-})+g_{1} \int dr
\cos(2\phi_{s+})\cdot \cos(2\phi_{s-})+ g_{2}\int dr
\sin(2\phi_{s-})\sin(2\phi_{s+})+ g_{\parallel c} \int dr
\cos(2\theta_{c-})\cdot \cos(2\theta_{s-})
$$
\end{widetext}
Here, we use the same convention for the Klein factors as in
\cite{tsuchiizu_2leg_firstorder}. Subscripts $1$ to $4$ have the
 standard \emph{g-ology} meaning and labels $a$ to $d$
refer to processes involving the $0$ and/or $\pi$ bands.
The two $g_{1d}$ terms, e.g, describe events where one right-
and one left- moving fermion, both belonging to the same  ( $0$ or $\pi$) band,
backscatter within that band.
 $g_1$ and $g_2$
correspond to the sum and to the difference of these ``1d''-type
processes respectively, and $g_2\neq 0$ when the $O$ atoms are
included. The $g_{4a}$ term has a non-zero conformal spin 
so that additional interactions
$G_{p(t)}\sim\cos(4\phi_{s-}(\theta_{s-}))$ are generated
during the flow. They are included in our
calculations. Since we are concerned with a priori incommensurate
values of $\delta$ we drop all umklapp terms. We renormalize the
couplings in (\ref{eq:kosinusy}) following the usual RG procedure,
where one integrates out high energy states. This sequence is
straightforward when the quadratic part (\ref{eq:Hbozon}) is
expressed in the $B_o$ basis, since one deals with simple gaussian
integrals but when we express $H^{NL}_{int}$ in the $B_o$ basis
this involves $P_{i}$ and $Q_{i}$ coefficients. Each RG step then
generates cross-terms in $H_0$, which implies a rotation of $B_0$
with respect to the $B_{+-}$ basis. It is thus important to
include the change in $S$ during the flow. After the $n^{th}$
iteration, we denote by ($\alpha_n,\beta_n$) the angles between
$B_o$ and $B_{+-}$ and by ${K^{(n)}}_{\lambda}$ the parameters in
the $B_o$ basis. We perform the $(n+1)^{th}$ RG step in the $B_o$
basis, which changes ${K^{(n)}}_{\lambda}$ (see (\ref{eq:Hbozon}))
and introduces cross-terms. We apply $S^{-1}(\alpha_n ,\beta_n)$
to $H_0$, which takes us back to the (fixed) $B_{+-}$ basis.
Finally we determine the new angles ($\alpha_{n+1}, \beta_{n+1}$)
which are required to make $H_{0}$ diagonal again, with new
parameters  ${K^{(n+1)}}_{\lambda}$.

Proceeding in incremental steps gives the additional
RG equations for the rotation of the $B_{o}$  basis
\begin{equation}\label{alpha}
  \frac{d\cot2\alpha(\beta)}{dl}=\frac{((dK_{1(3)}-dK_{2(4)})\tan4\alpha+dB_{12(34)})}{K_{1(3)}-K_{2(4)}}\cdot
  dl^{-1}
\end{equation}
The equations for the off-diagonal terms $dB_{12(34)}$ are
\begin{eqnarray}\label{B12B34}
\frac{dB_{12}}{dl}&=&P_{1}Q_{1}((g_{1a}^{2}+g_{\parallel
c}^{2}+G_{t}^{2})-K_{1}K_{2}(g_{1a}^{2}+g_{1c}^{2}+ \nonumber
\\&& +g_{2c}^{2}+G_{p}^{2}))-K_{1}K_{2}h(P_{1})g_{1}g_{2} \nonumber
\\&& \frac{dB_{34}}{dl}=P_{2}Q_{2}(g_{1c}^{2}+g_{2c}^{2}+ g_{\parallel c}^{2})
\end{eqnarray}
with $h(P_{1})=((P_{1}Q_{1})^{2}+0.25(P_{1}^{2}-Q_{1}^{2}))^{-1}$. Details of the RG equations for the various $g$, $K$ and for the ratio of the Fermi velocities in the $0$ and $\pi$ bands will be given in a forthcoming publication \cite{chudzinski_ladder_long}.

Using the above equations we establish the phase diagram for the ladder.
In agreement with \cite{balents_2ch} we find that
$\tilde{\alpha}=\frac{V_{Fo}+V_{F\pi}}{2 V_{Fo}} $ 
(a ratio of Fermi velocities in the $0$ and $\pi$ bands)
controls the behavior of the differential equation system.
When
$\frac{t_{\bot}}{t}$ is constant,
$\tilde{\alpha}$ depends only on $\delta$.
Two main factors may significantly affect the phase
diagram that was predicted for two leg Hubbard ladders
with a single orbital per site: one is the
asymmetry in the $g$ terms due to the fact that
the projections of the $Cu$ and $O$ orbitals onto
the $0$ and $\pi$ bands have  unequal amplitudes and one
is the influence of the extra parameters $U_{O}$, $V_{Cu-O}$ and $t_{pp}$.
We first investigate the impact of the asymmetry, so we set $U_{O}=V_{Cu-O}=t_{pp}=0$ and we choose
small initial values for $U_{Cu}$ (in the range $10^{-6}-10^{-1}$).

{\it (a) Small doping range. } For small $\delta$
($\tilde{\alpha}$),  $\cot2\alpha \rightarrow 0$ and $\cot2\beta
\rightarrow 0$ as the flow converges towards the fixed point, thus
$B_o\to B_{+-}$. In this case, $ g_{2},g_{4a},G_{p},G_{t}$ are
irrelevant while $ \theta_{c-} $ and $ \phi_{s+} $ are ordered ($
\theta_{c-}=0 $, $\phi_{s+}=0 $ mod $ 2\pi$). This is the
\emph{C1S0} phase \cite{balents_2ch} where only the total (+)
charge mode is massless. For the $s-$ (spin-transverse) mode,
terms involving the canonically conjugated fields $ \phi_{s-}$ and
$ \theta_{s-}$ are relevant and competing. d-type superconducting
fluctuations (SCd) dominate if $\phi_{s-}$ is locked at $0$, while
OAF is favored if $\theta_{s-}=0$. Here, SCd is always more stable
for repulsive $U_{Cu}$. This property holds only for
$\delta<\delta_{c1}=0.2$, where the spin and mass gaps go to zero.

{\it (b) Large doping range. }
For $\delta>\delta_{c2}=0.28$,
$\cot2\alpha$ and $\cot2\beta$ $ ~\rightarrow~\infty$ (with opposite signs), so $B_o \to B_{0\pi}$.
In this regime only $ g_{1}\simeq -g_{2}$ are relevant and they
lead to a state
with one massive spin mode (in the $0$ band). This is the $C2S1$
phase.
The slowest decay of correlations is observed for the CDW operator
in the $0$ band. Fluctuations in the $\pi$ band favor a SDW state
(when logarithmic corrections due to
a marginal operator proportional to $g_{1}+g_{2}>0$ are included) but they are subdominant.
If $\delta$ is just above $\delta_{c2}$,
$g_{1}$ and $ g_{2}$ increase
very slowly during the flow and one needs to choose larger values for the bare $U_{Cu}$  (still much smaller than $t$)
to reach the asymptotic regime with a gap in the spin
mode.
In contrast with the case of a single orbital per site,
the $C2S1$ phase is stable, even for dopings such that $E_F$ is close to
the bottom of the $\pi$ band where $\tilde{\alpha}$ is very large
(in that limit, we cannot linearize the energy spectrum, but we use 
diagrammatic techniques \cite{balents_2ch}).
This comes from the fact that for unit cells with $Cu$ atoms only, $g_{2}$
is accidentally equal to zero. When $O$ atoms are included  (or when
$V_{Cu-O}\neq 0$) the initial $g_{2}$ is non-zero and $g_2$ is
always relevant.
The nature of this \emph{C2S1} phase is discussed in the next paragraph.

{\it (c) Massless regime in the ($\delta_{c1},\;\delta_{c2}$) range.}
As $\delta$ approaches the critical end points $\delta_{c1}$ and
$\delta_{c2}$ respectively from below (in the \emph{C1S0} phase) and from above (in the \emph{C2S1} phase),
gaps in the spin and/or in the charge sectors go to zero.
All spin and charge modes are massless in the entire range of dopings
$\delta\in(0.2;0.28)$.
$\frac{d\alpha}{dl}$ and $\frac{d\beta}{dl}$
are very large and the fixed point values of $\beta$ ($\alpha$)
just below and just above $\delta_{c1}$ ($\delta_{c2}$) are significantly different.
So we approach the critical
points from the massive phases at both ends; we discard couplings
which flow to zero and thus obtain a simpler
set of equations. Next
we single out terms in (\ref{alpha}-\ref{B12B34}) which produce
large values of the derivatives in this range
and determine the fixed point value of $\cot2\alpha(\beta)$.
It allows us to write down a minimal set of
differential equations for the couplings and to determine those which are
relevant in the doping range ($\delta_{c1},\;\delta_{c2}$).
We first consider dopings close to $\delta_{c2}$. This
corresponds to an initial value of $\cot2\alpha$ equal to one.
The signs of $(dK_{1}-dK_{2})$ and $dB_{12}$ are the same and
positive whereas the sign of $(K_{1}-K_{2})$ is negative so that,
according to (\ref{alpha}), $|\cot2\alpha|$ decreases to zero
below $\delta_{c2}$ while above it increases to infinity.
Below $\delta_{c2}$, $g_{1}$ and $ g_{2}$ are not relevant and the
system flows to the \emph{C2S2} phase while above they are
relevant, leading to the \emph{C2S1} phase.
For $\delta$ close to $\delta_{c1}=0.2$,
$(dK_{3}-dK_{4})$ and $dB_{34}$ have opposite signs. Depending on
which of the two terms dominates, $|\cot2\beta|$ goes to zero or
to infinity. At $\delta_{c1}$ they are exactly equal.
$|\cot2\beta|\rightarrow\infty$, for $\delta >\delta_{c1}$, but, since
$0.5<K_{4}<1$, one finds that all
interband couplings 
(as well as higher order $\cos(b\phi_{c})$ terms with $b=4,6,..$) 
are irrelevant. 
%
%
For
$\delta\in(0.2;0.28)$, all interaction terms are irrelevant
and $B_{0\pi}$ ($B_{+-}$) is the eigenbasis for the charge (spin) modes.
A numerical solution of the full set of RG
equations confirms this statement. The existence of this massless
regime is essential to maintain spin rotational symmetry in this doping range.
The nature of the \emph{C2S2} phase can be determined
in the framework of the Luttinger liquid description.
In that phase, $K_{4}$ ($K_{3}$) which corresponds to the $0$
($\pi$) band is smaller than (around) one.
Dominant fluctuations occur in the $0$ band, and this case is equivalent
to treating a single chain problem.
The only marginal couplings
are $g_1$ and $g_2$ ($g_{1}>g_{2}$). Including logarithmic corrections allows us to
identify the slowest decaying correlation function
and we find that in the  ($\delta_{c1},\;\delta_{c2}$) doping range, a SDW in the
$0$ band ($SDW$(o)) is favored.
In the \emph{C2S1} state, $g_1\simeq -g_2<0$ are relevant which
gives a gap in the spin sector of the $0$ mode. In that regime,
fluctuations in the $0$ band dominate, and one finds that the
$CDW$(o) state is the slowest decaying one.

Next, we ``turn on'' the parameters $U_{O}$, $V_{Cu-O}$ and $t_{pp}$
and we examine their impact
on the phase diagram. In the doping range covered by case (a), SCd
becomes less dominant over OAF when we increase the (positive) bare
value of $U_{O}$ or
$V_{Cu-O}$ at fixed $t_{pp}$ but it is still the phase with the
lowest free energy. One would need to assume a very large
attractive bare $V_{Cu-O}$
to cause a
transition \cite{lee_marston_CuO} to an s-type SC phase
($\phi_{s-}=0$, $\phi_{s+}=0$, $\theta_{c-}=\pi/2$)
\cite{tsuchiizu_2leg_firstorder}
which persists
even as $E_F$ approaches the bottom of the $\pi$ band.
As far as case (b) is concerned, we observe a reduction in the
size of the gap for positive $U_{O}$ and $V_{Cu-O}$, while for
very large attractive $V_{Cu-O}$ the s-SC phase re-enters. In case (c),
increasing $V_{Cu-O}$ has little effect on $\delta_{c2}$ but it
shifts $\delta_{c1}$ to higher values. An unphysically large ratio
$ V_{Cu-O}/U_{Cu}\approx 5$ would be required to suppress the
massless phase and to observe a reentrant \emph{C1S0} phase with
superconducting fluctuations so that in the relevant case $
V_{Cu-O}< U_{Cu}$ the intermediate massless phase does exist.

The interoxygen hopping has a more significant effect. Increasing the value of
$t_{pp}$ causes a concomitant decrease of $\delta_{c1}$ and
$\delta_{c2}$. For $t_{pp}=0.5$ --a value pertaining to $Cu-O$ ladders
\cite{Muller-Rice_LDA}-- their values are about half that
quoted for $t_{pp}=0$. If $t_{pp}>t_{pp}^{min}$, a new
phase is favored when $\delta>\delta_{c1}$. This state has both
orbital current and $DW$ fluctuations ($DW\equiv SDW\; (CDW)$ for
the $C2S2$ ($C2S1$) regime) and it shows similarities with one of
the patterns advocated by Varma \cite{CVarma_orbitalcurrents} (see
Fig. \ref{fig:structure}b). This current phase is an eigenstate of
the $B_{0\pi}$ basis (it is invariant under the exchange of the
two legs) in contrast with the other Varma pattern or with the
usual OAF. The pattern has an incommensurate spatial periodicity
$\sim k_{F0}^{-1}$. The amplitude of this order parameter is a sum
of current operators between links of the $Cu$-$O$ loops, of the
form $t_{ij}Im(\lambda_{i 0}^{*}\lambda_{j 0})$,
where $t_{ij}$ is the hopping parameter from site $i$ to site $j$ within the same unit cell
and $\lambda_{i 0}$ is the overlap between the ($Cu$ or $O$) wavefunction at site $i$ and the $0$ band eigenfunction.
These quantities are of order one \cite{chudzinski_ladder_long} and
change by only a few percent when $\delta$ increases from
$\delta_{c1}$ to the bottom of the band.
Due to current
conservation, the weakest link between atoms determines
the maximal current, and we find that, for
$t_{pp}^{min}\approx 0.3$,  the \emph{"Varma"}
state dominates the \emph{DW(o)}. The presence of the $O$
sites insures that
$Im(\lambda_{i\alpha}^{*}\lambda_{j\alpha})\neq 0$
($\alpha=0,\pi$);
Otherwise,
the current operator between Cu atoms has the usual interband
form: $c_{o\sigma}^{\dagger}c_{\pi\sigma}$. 

Our predictions could be tested by performing NMR measurements
on the $Sr_{14-x}Ca_xCu_{24}O_{41}$ compound \cite{dagotto_ladder_review} where the hole content can be somewhat varied, as
we find very different responses of the the spin modes for the
$Cu$ and $O$ sites \cite{chudzinski_ladder_long}.
\begin{figure}
 \begin{center}
  \includegraphics[width=\figwidth]{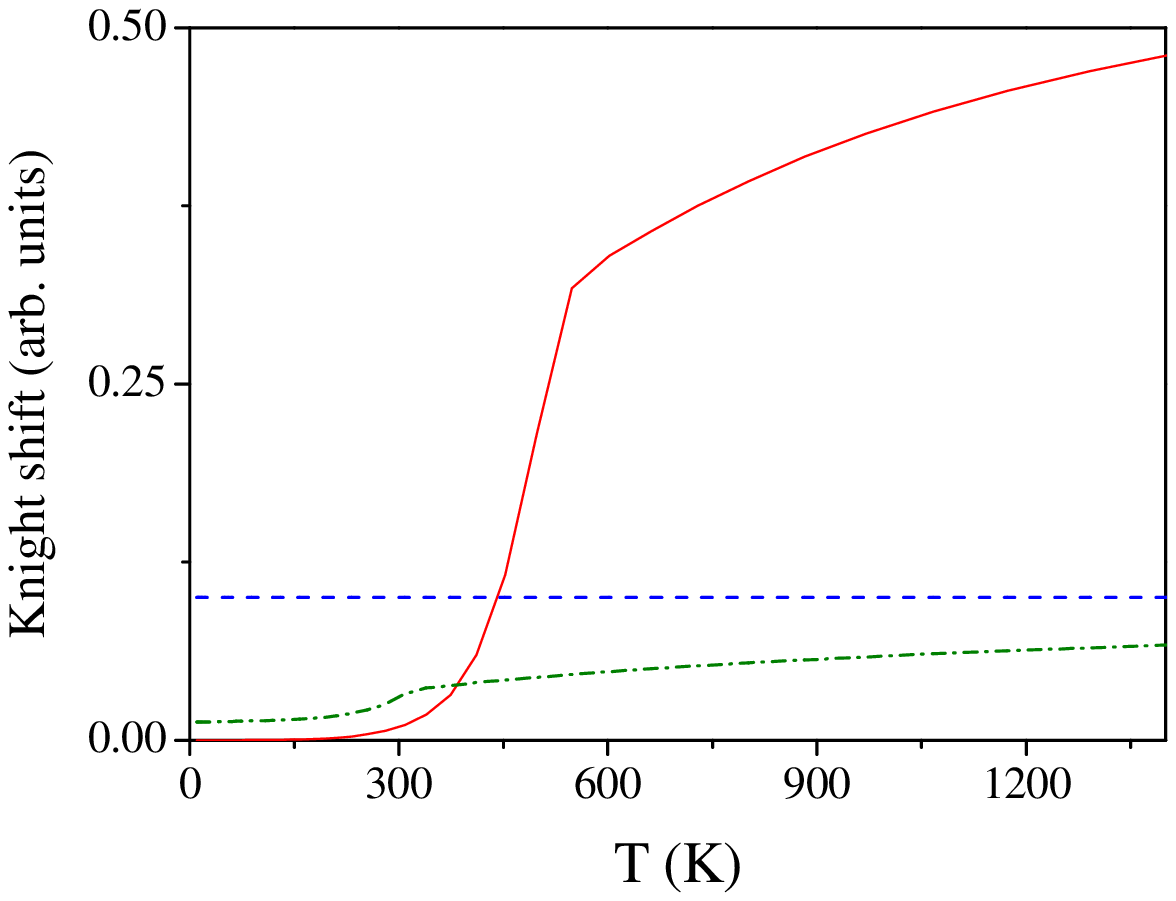}
 \end{center}
 \caption{(Color online) Knight shift on the outer $O$ sites calculated for different phases:
 solid line \emph{C1S0}, dashed line \emph{C2S2},  dash dotted line \emph{C2S1}.
 \label{fig:Kphases}}
\end{figure}
Knight shifts $K$ and relaxation rates $T_{1}^{-1}$ can be
evaluated in all three regimes. For low doping we find an
activated behavior of $K$ (and also $T_{1}^{-1}$) and $K(T=0)=0$;
for high doping the temperature dependance is similar but
$K(T=0)\neq 0$; finally, at intermediate dopings the usual high
temperature behavior $K\sim T^{0}$ (saturation to the LL value)
extends to $T\rightarrow 0$. For instance, \fref{fig:Kphases} shows the
Knight shift predicted for the "outer" $O$ sites (i.e interladder bridges)

In conclusion, we have developed a new RG method to handle correlation effects
in the weak coupling limit for
two leg Hubbard ladders at generic filling, when oxygen atoms are included in the unit cell.
We have found a ground state phase diagram where the
\emph{C1S0} and \emph{C2S1} phases are present at small and large dopings,
as for the single orbital ladder, but also a new intermediate phase
\emph{C2S2} which is completely massless. Dominant fluctuations in
the \emph{C2S2} and  \emph{C2S1} states correspond to orbital
currents preserving the mirror symmetry of the ladder structure on
top of a SDW(o) and CDW(o) respectively. The stability of this new
phase to an interladder coupling and/or to large values of the
bare $U$ are under current investigation.

{\it Acknowledments.}
The authors are indebted to T. Becker and to C.M. Varma.
This work was supported in part by the Swiss NSF under MaNEP and Division II and by an ESRT Marie Curie fellowship.

\bibliography{totphys,ladder3b}

\end{document}